\documentclass[11pt,english]{article}
\usepackage[letterpaper]{geometry}
\usepackage[T1]{fontenc}
\usepackage[latin9]{inputenc}
\usepackage{longtable}
\usepackage{tikz}
\usetikzlibrary{positioning}
\usepackage[onehalfspacing]{setspace}
\usepackage{amsmath,amsfonts,bbm}
\usepackage{amsthm}
\usepackage{graphicx}
\usepackage[authoryear]{natbib}
\PassOptionsToPackage{normalem}{ulem}
\usepackage{ulem} 
\usepackage{multirow}
\usepackage{booktabs,makecell} 
\usepackage[labelformat=simple]{subcaption} 
\usepackage[labelfont=bf,textfont=bf]{caption} 
\usepackage{color}
\definecolor{bl}{rgb}{0,0.1,0.5}
\definecolor{dkblue}{rgb}{0,0,0.65}
\definecolor{darkgreen}{rgb}{0,0.5,0}
\usepackage{tabularx}
\newtheorem{thm}{Theorem}

\newtheorem{ass}{Assumption}

\usepackage[bookmarks=true,bookmarksopen=true,bookmarksopenlevel=0,pdfview=fit,
	pdffitwindow=true,colorlinks=true,urlcolor=dkblue,linkcolor=dkblue,citecolor=dkblue]{hyperref}
	
\usepackage[colorinlistoftodos]{todonotes}

\begin{document}
\addtocounter{footnote}{2}
\title{LATE for History\thanks{%
Prepared for the {\em Handbook of Historical Economics}, by Alberto Bisin and Giovanni Federico, eds, Elsevier North-Holland, forthcoming 2021. We thank Thierry Verdier for countless exchanges on these topics over the years. We also thank Jess Benhabib, Giovanni Federico, Atsushi Inoue, Franco Peracchi, and Pedro Sant'Anna for helpful comments} }
\author{Alberto Bisin\thanks{%
Department of Economics, New York University, and NBER, alberto.bisin@nyu.edu} \and Andrea Moro\thanks{Department of Economics, Vanderbilt University, andrea@andreamoro.net}
}
\maketitle

\begin{abstract}
In Historical Economics, Persistence studies document the persistence of
some historical phenomenon or leverage this persistence to identify causal
relationships of interest in the present. In this chapter, we analyze the
implications of allowing for \emph{heterogeneous treatment effects} in these
studies. We delineate their common empirical structure, argue that heterogeneous
treatment effects are likely in their context, and propose minimal abstract
models that help interpret results and guide the development of empirical
strategies to uncover the mechanisms generating the effects.
\end{abstract}

\newcommand{\kwd}[1]{ #1 }
Keywords: 
\kwd{Local average treatment effects}
\kwd{Persistence studies}
\kwd{Instrumental variables}
\kwd{Identification}
\kwd{Institutions}
\kwd{Culture}



\section{Introduction}

\newcommand{\xch}[2]{#1}
Many studies in Historical Economics document the persistence of some historical
phenomenon, while others leverage this persistence to identify causal relationships
of interest in the present. These are generally referred to as
\emph{Persistence studies} \citep[in this book]{Voth}. In this chapter, we analyze the
implications of allowing for \emph{heterogeneous treatment effects} in these
studies\xch{, we}{, We} delineate their common empirical structure, argue that heterogeneous
treatment effects are likely in their context, and propose simple abstract
models that help interpret results and guide the development of empirical
strategies to uncover the mechanisms generating the effects.

Persistence studies focus empirically on the effects of a treatment variable
in the present, assuming its persistence from the historical past. Consider
as an illustration the effects of cultural norms or institutions, e.g.,
on economic development. The adoption of cultural norms or the process
of institutional change can be viewed, using the language of the causal
inference literature, as taking-up treatment in the past. High-quality
norms or institutions may then persist over time and thereby realize their
effects on economic development in the present. An exogenous historical
factor may be available that directly affects the treatment variable and
can be exploited as an instrumental variable to identify a causal effect
of norms or institutions on development.

The adoption of norms or institutions may be correlated with the returns
in terms of economic development. If these returns are heterogeneous across
countries - that is, if treatment effects are heterogeneous - countries
with higher returns may be more likely to have adopted higher quality institutions
or norms.\footnote{This is often labeled a ``selection on returns'' or ``selection
on gains,'' see \cite{heckman2010testing}.} Heterogeneous treatments may
then change the interpretation of the identified relationship between norms
or institutions and economic development. For example, if higher values
of the instrument induce countries with relatively higher returns to adopt
higher quality norms or institutions, then the instrumental variable procedure
identifies the effects produced only in high-return countries, possibly
overstating the average returns of all countries. Similarly, the instrument
could activate institutional and cultural changes over time, interacting
with successive independent historical phenomena, inducing a special selection
of heterogeneous treatment effects. Finally, even if the impact of institutions
and norms in the present is homogeneous, the persistence of institutional
and cultural changes in the historical past activated by treatment can
be heterogeneous, affecting the interpretation of the nature of persistence.

When treatment effects are heterogeneous, therefore, the causal arguments
remain generally unaffected; but the interpretation of estimated coefficients
may pose new empirical questions to analyze further the mechanisms underlying
the identified causal relationships. More specifically, when treatment
effects are heterogeneous, the research design identifies a \emph{Local
Average Treatment Effect} (LATE), rather than the Average (subject-level)
Treatment Effect (ATE); see \cite{angrist1995identification}. In the investigation
of the effects of cultural norms or institutions, for example, the \emph{local}
effect identified by the instrumental variable procedure is the effect
in countries that take-up treatment (higher quality institutions or norms)
only when instrumented.\footnote{In environments where this
\emph{essential heterogeneity} is present \cite{heckman-vytlacil-2005} show
that the IV procedure identifies a weighted average of the returns and
characterize the weights in terms of the relationship between the IV and
the returns. In the Appendix we introduce a simple, informal primer on
the distinction between LATE and ATE in the context of labor economics,
to level the field between historians, economists, and other social scientists
who might not be uniformly attuned to these concepts.}

Heterogeneous effects are bound to be important in Persistence studies.
In fact, the persistent variable in Persistence studies is typically a
cultural (or genetic) trait or an institutional feature or norm.\footnote{Recent
examples have exploited variation in a wide range of historical variables:
from colonial settlers' mortality to the self-governance of Italian cities,
the adoption of the plow in agriculture, medieval pogroms, the size of
the slave trade from Africa. Some of these studies have identified causal
relationships between phenomena of interest in the present, for example,
the effect of institutional quality or civic capital on economic performance.
Others have documented the persistence of various dimensions of institutional
and cultural characteristics, including trust, gender attitudes, and anti-semitism.
Persistence studies are a sizeable component of Historical Economics, about
$10\%$, according to \citet[in this book]{cionitwo}'s classification; see their Table
2 and 4.} Because these variables are often defined by such sweeping concepts
such as institution quality, cultural, or civic norms, they are likely affected
by several underlying heterogeneous mechanisms. Furthermore, because treatment
is generally taken-up in the historical past, various complex and heterogeneous
dynamical processes may affect it, intervening in the determination of
the objective of the analysis in the present.

Since heterogeneous treatment effects do not generally affect the causal
identification argument, in this chapter we shy away from discussing the
validity of the data and econometric procedures adopted in these studies.\footnote{Persistence
studies have been the subject of great scrutiny. The credibility of the
causal parameter estimates stands on the reliability of historical data,
assumptions about the econometric model's error structures, and, in most
cases, on the adoption of a valid instrumental variable (\cite{CASEY2021102586}). Because most of
these studies exploit variation across geographic dimensions, spatial correlations
of residuals can be a severe econometric issue; see
\cite{kelly2019standard} and \citet[in this book]{Voth}.} We focus instead
on the interpretation of the estimated coefficients under heterogeneous
treatment effects. While the identification of causal effects in Persistence
studies has produced significant first-order results, it has also highlighted
how little we know about the mechanisms driving these effects. The paucity
of data in historical contexts makes mechanisms hard to identify both with
a quasi-experimental design and with a structural econometrics approach.
For the same reason, it is hard to identify the distribution of treatment
heterogeneity.

To guide our understanding about what moments of this treatment parameters
distribution the estimates identify, it is therefore important to study
the relationship between the instrument and treatment and how the mechanism
responsible for the persistence of treatment over time correlates with
the values of treatment effects. Explicit models of these relationships
and mechanisms, in the context of the specific empirical analysis, help
to clarify the interpretation of the identified causal effects and help
the formulation of interesting new sets of questions, which can be addressed
empirically, possibly with new data. To this end, we develop simple abstract
models linking treatment take-up and treatment's persistence over history
with treatment effects. These models delineate the underlying causal relationships
and the role of the research design in the identification problem. Explicit
models of these relationships and mechanisms represent the outcome of political
equilibrium processes or the aggregation of individuals' relevant behavioral
choices. This is typically the case, in particular, when treatment involves
institutional change or change in cultural attitudes and traits as just
described. To illustrate the role of these models in the abstract, without
imposing a fully developed structure (which would be necessarily context
dependent), we will construct minimal reduced-form models of treatment
take-up
 without being explicit about their behavioral and equilibrium micro-foundations.
These models provide us with interpretations of the estimated parameters,
opening new empirical questions, possibly with new data.

We proceed, in the next section, by formalizing Persistence studies' common
structure. In Section~\ref{sec:current} we apply this framework to persistence
studies whose main goal is to identify the causal relationship between
variables in the present. These studies exploit the persistence of a variable in the
historical past to provide an instrument. In Section~\ref{sec:purepersistence}, we focus instead on a set of persistence studies
that are directly interested in investigating the persistence of historical
variables.

\section{Persistence studies}
\label{sec:theory}

In this section, we introduce the issue of identification of causal relationships
in Persistence studies. We delineate their common empirical structure by
constructing a simple formal framework that encompasses most papers in
the literature.\footnote{For a book-length treatment of causal analysis
in econometrics, see \cite{angrist2008mostly,angrist2014mastering}. For
a more abstract approach to causality, see
\cite{pearl2009causality,pearl2016causal}.}

\subsection{Empirical model}
\label{em}

Let any variables measured in present time be indexed by $t$ and any historical
variables by $t-h$, where $h$ is the historical lag considered. Let
$i=1,2, \ldots , N $ index the cross-section in the data, typically defined
along political or geographic dimensions: countries, cities, ethnic groups,
etc\ldots \
Consider the following cross-sectional relationship between variables at present time $t$:
%
\begin{equation}
y_{t}=\alpha + \beta x_{t} +u_{t},
\label{EmpirMod}
\end{equation}
where $y_{t}, x_{t}, u_{t}$ are $N-$dimensional vectors indexed by location
$i=1,2, \ldots , N $; the
parameter $\beta $ is also an $N-$dimensional vector: distinct
$\beta _{i}$, across locations $i=1,2, \ldots , N$, represent the heterogeneous
effects which are the focus of our analysis.\footnote{We abuse notation by not distinguishing the
random variables in the population from their sample realizations. Also, in our notation, products between vectors are to be intended as Hadamard products; so that, in Equation \ref{EmpirMod},  $\beta x_t =\left[ \beta_{i}x_{i,t}\right]_{i=1}^N$.}  
The explanatory variable
$x_{t}$ is generally endogenous, e.g., because of a common factor affecting
both $x_{t}$ and $y_{t}$ or because of two-way causation between
$y_{t}$ and $x_{t}$.

History enters the empirical model through the underlying (unobservable)
historical dynamics of the explanatory variable $x_{t}$, governed by a
stochastic process, $\{x_{\tau}\}_{\tau \in T}$, where $T$ denotes the
historical sequence of time until the present
$\{t-h, \ldots , \tau , \ldots t\}$. We model the persistence of the process
$\{x_{\tau}\}$ assuming
\begin{equation}
cov( x_{t-h}, x_{t})=\rho ;
\label{rho}
\end{equation}
where $\rho $ is also an $N-$dimensional vector, allowing for heterogeneity
of persistence across locations $i\xch{= 1,}{=, 1,} 2, \ldots , N$ (for simplicity we assume that $\rho$ does not depend on $h$ or $t$).

\newcommand{\reftext}[1]{#1}
The econometrician observes an instrument for $x_{t}$ that we will assume to be valid throughout the chapter, in the sense defined below (Assumption \ref{ass:currentrel}). The characterizing
feature of Persistence studies is that the instrument, an $N-$di\-men\-sional
vector $z_{t-h}$, is realized in the historical past, at $t-h$. The empirical
structure underlying the IV strategy depends crucially on the historical
persistence of the instrumented variable, $x_{t}$. The instrument
$z_{t-h}$ is assumed to affect causally and directly $x_{t-h}$ and hence
indirectly $x_{t}$ through the persistence of the process
$\{x_{\tau}\}_{\tau \in T}$ in history. This structure does not require
the econometrician to observe the realizations of the stochastic process
at any time other than $t$; in particular, $x_{t-h}$ is generally not observable
to the econometrician. However, if a proxy $p_{t-h}$ for $x_{t-h}$ is observable,
$z_{t-h}$ can be an instrument for $p_{t-h}$, even though $p_{t-h}$ might
be endogenous with respect to $x_{t-h}$. \reftext{Fig.~\ref{fig:EmpirMod1}} illustrates
the relationship between variables in this empirical model.%
\footnote{It is difficult to represent models with endogenous, equilibrium
relationships, or to represent the distinction between LATE and ATE, using
Directed Acyclical Graphs (see \cite{Imbens_2019}). This chart and the
next ones are meant to illustrate the relationships between variables in
the various empirical models, not as a tool to discuss the causal identification
designs, which we study analytically in the text.}\looseness=1

\begin{figure}[t]
%
\centering{}
\begin{tikzpicture}%
[observed/.style={circle,draw=black,fill=white,minimum size=11mm},
         unobs/.style={circle,draw=black,fill=white,minimum size=11mm,dashed},
         timetext/.style={draw=white,fill=white},
         node distance=2cm,
         on grid,
         >=latex
         ]
          \node[observed] (x) {$x_{t}$};
          \node[observed,right= of x] (y) {$y_{t}$};
          \node[unobs,below= of x] (xtau) {$x_{\tau}$};
          \node[unobs,below= of xtau] (xth) {$x_{t-h}$};
          \node[observed,left= of xth] (z) {$z_{t-h}$};
          \node[observed,right= of xth] (p) {$p_{t-h}$};
          \node[timetext,right= 10mm of y,anchor=west] (pres) {present ($t$)};
 		  \node[timetext,right = 30mm of xtau, anchor=west] (past) {past ($\tau $)};
 		 \node[timetext,right= 10mm of p,anchor=west] {historical past ($t-h$)};
          \node[timetext,above right=.25cm and 1cm  of x] {${\color{red} \beta}$};
          \node[timetext,above left=1cm and .3cm  of xtau] {${\color{red} \rho}$};
          \node[timetext,below left=1cm and .3cm  of xtau] {${\color{red} \rho}$};
          \draw [<->]
            (x) edge (y)
            (p) edge (xth);
          \draw [->]
             (z) edge (xth)
            (xth) edge (xtau)
            (xtau) edge (x)
            (z) .. controls +(down:15mm) and +(down:15mm) ..  (p);
        \end{tikzpicture}
%
\caption{Persistence studies: general case.}
\label{fig:EmpirMod1}
\caption*{\normalfont \small
\textit{Note}: \normalfont \small \xch{Circles}{circles} indicate variables observed
by the investigator. Dashed circles indicate unobserved variables. Solid
arrows indicate directions of causality. Double arrows indicate endogeneity
or any other factor preventing the identification of a causal effect (omitted
variables, selection bias, etc...).}
\end{figure}
The examples from the literature that we study in the rest of the paper
focus on two separate objectives of the empirical analysis. One is to estimate
some moment of the cross-section of treatment effects,
$\beta _{i}$, or another relevant feature of their distribution. When data on a proxy $p_{t-h}$ for $x_{t-h}$ are available,
another objective of the analysis is to estimate some relevant feature of
the distribution of the cross-section of persistence effects, $\rho _{i}$.

\subsection{Heterogeneity and LATE effects}

Allowing for heterogeneous treatment effects might have important implications
with respect to how these effects, as they are identified in Persistence
studies, are interpreted. Consider the empirical model delineated in Section~\ref{em} and represented in \reftext{Fig.~\ref{fig:EmpirMod1}}. Assume that either
the relationship between $x_{t}$ and $y_{t}$ and/or the persistence effects
from $x_{t-h}$ to $x_{t}$ are heterogeneous; that is,
$\beta , \rho $ are heterogeneous across locations
$i=1,2, \dots , N$. As long as $z_{t-h}$ is a valid instrument for
$x_{t}$ or for $p_{t-h}$, the IV procedure identifies a causal effect;
that is, the causal arguments which are the objective of Persistence studies
remain unaffected. But consider the case in which the mechanism inducing
location $i$ to take-up treatment after being instrumented, or to maintain
the treatment over time, depends on the values of $\beta $ and
$\rho $ assumed by that location. In this case, heterogeneity matters for
the interpretation of the identified causal effect. For instance, if locations
taking-up treatment have on average high $\beta $, e.g., because a high
$\beta $ lowers the cost of treatment, then the IV procedure will tend
to identify a high $\beta $; that is, the \emph{local} effect of treatment
for those locations in fact induced by the instrument to adopting treatment
(the \emph{complier} locations in the jargon of LATE studies; see later).

To push the empirical analysis of Persistence studies forward it is therefore
important to study the relationship between the instrument $z_{t-h}$ and
treatment $x_{t}$ and how this relationship can be filtered through the
values of $\beta $ and/or $\rho $. Similarly, it is important to study
how the mechanism responsible for the persistence of treatment over time,
from $x_{t-h}$ to $x_{t}$, correlates with the values of $\beta $ and/or
$\rho $. Explicit models of these relationships and mechanisms, in the
context of the specific empirical analysis, help clarifying the interpretation
of the identified causal effects and help the formulation of interesting
new sets of questions, which can be addressed empirically, possibly with
new data. To this end, we develop below minimal abstract models linking
treatment take-up in a location, and the treatment's persistence over history
with the location's value of $\beta $ and/or $\rho $.

Importantly, the mechanisms leading to treatment take-up and its persistence
over history cannot generally be represented by behavioral choice
problems because the units of observation in most of these studies are
countries, cities, ethnic groups. Consequently, whether treatment is adopted
or not, and whether it persists over history, rests on the outcome of political
equilibrium processes or on the aggregation of the relevants behavioral
choices of individuals. This is typically the case when treatment involves
institutional change or change in characteristic cultural attitudes and
traits. For the sake of simplicity and abstraction, our analysis will be
limited to reduced-form models of treatment take-up, without being explicit
about their equilibrium micro-foundations. Furthermore, since the context
determines the appropriate modeling assumptions, we will illustrate our
approach in the context of several classic Persistence studies.

\begin{figure}[t]
\begin{tabularx}{\textwidth}{ccc}
{\centering
\begin{tikzpicture}%
             [observed/.style={circle,draw=black,fill=white,minimum size=11mm},
              unobs/.style={circle,draw=black,fill=white,minimum size=11mm,dashed},
              node distance=2cm,
              on grid,
              >=latex
            ]%
              \node[observed,red,fill=none] (x) {$x_{t}$};
              \node[observed,right= of x,red,fill=none] (y) {$y_{t}$};
              \node[unobs,below= of x] (xtau) {$x_{\tau}$};
              \node[unobs,below= of xtau] (xth) {$x_{t-h}$};
              \node[observed,left= of xth] (z) {$z_{t-h}$};
              \node[observed,right= of xth,white] (p) {$p_{t-h}$};
          \node[above right=.25cm and 1cm  of x] {${\color{red} \beta}$};
              \draw [<->,red]
                (x) edge (y);
              \draw [->,black];
              \draw [->]
                (xth) edge (xtau)
                (xtau) edge (x)
                (z) edge (xth);
             \draw [white] (z) .. controls +(down:15mm) and +(down:15mm) ..  (p);
        \end{tikzpicture}}
 &
{\centering\begin{tikzpicture}%
         [timetext/.style={},
           node distance=2cm,
         on grid,
         >=latex
         ]
          \node[timetext] (t) {present};
          \node[timetext,below= of t] (tau) {past};
          \node[timetext,below= of tau] (th) {historical past};
        \draw [white] (th) .. controls +(down:15mm) and +(down:15mm) ..  (th);
        \end{tikzpicture}}
&
{\centering\begin{tikzpicture}%
             [observed/.style={circle,draw=black,fill=white,minimum size=11mm},
              unobs/.style={circle,draw=black,fill=white,minimum size=11mm,dashed},
              node distance=2cm,
              on grid,
              >=latex
            ]%
              \node[observed,red,fill=none] (x) {$x_{t}$};
              \node[unobs,below= of x,red,fill=none] (xtau) {$x_{\tau}$};
              \node[unobs,below= of xtau,red,fill=none] (xth) {$x_{t-h}$};
              \node[observed,left= of xth] (z) {$z_{t-h}$};
              \node[observed,right= of xth] (p) {$p_{t-h}$};
         \node[above left=1cm and .3cm  of xtau] {${\color{red} \rho}$};
          \node[below left=1cm and .3cm  of xtau] {${\color{red} \rho}$};

            \draw [->,red]
                (xth) edge (xtau)
                (xtau) edge (x);
             \draw [->] (z) .. controls +(down:15mm) and +(down:15mm) ..  (p);
              \draw [<->] (p) edge (xth);
               \draw [->,dashed]    (z) edge (xth);

        \end{tikzpicture}}
\\ {(a) Current variables}&& {(b) Pure persistence}
\end{tabularx}
\caption{Persistence studies: empirical models.}
\caption*{\normalfont \small
\textit{Note}: \xch{Circles}{circles} indicate variables observed
by the investigator. Dashed circles indicate unobserved variables. Solid
arrows indicate directions of causality. Double arrows indicate endogeneity
or any other factor preventing the identification of a causal effect (omitted
variables, selection bias, etc...).\ Dashed arrows indicate potential causal
links. Highlighted in \xch{red (gray in print version)}{red}  are the relationships of interest.}
\label{fig:EmpirMod2}
\end{figure}

In the next two sections\xch{, we}{, We} introduce two empirical models, special cases
of the model introduced above, to highlight different strands of the literature.\footnote{This
categorization is related to the distinction between \emph{apples-on-apples}
and \emph{apples-on-oranges} models by \citet[in this book]{Voth}.} In
Section~\ref{sec:current} we consider first the special case of the empirical
model in Section~\ref{em} in which the econometrician aims at uncovering
a causal relationship between current variables. The instrument in the
historical past and the persistence of the independent variable are, in
a sense, means to this end. The relationships between variables in this
class of models are represented in \reftext{Fig.~\ref{fig:EmpirMod2}}(a). As an illustration
of this empirical model it will be useful to consider two classic papers
among the most-well known in the field: \cite{acemoglu2001colonial} on
the colonial origins of economic development, and
\cite{ashraf-galor-2013} on the effect of human genetic diversity on development.
These papers provide distinct causal explanations of comparative economic
development (affected by different dependent variables).
\cite{acemoglu2001colonial} focus on the effect of the quality of institutions;
specifically, the level of protection of property rights.
\cite{ashraf-galor-2013} study how human genetic diversity affects economic
development through the contrasting effect of social conflict and innovation
or creativity.

Next, in Section~\ref{sec:purepersistence} we consider the special case
of the empirical model in Section~\ref{em} in which the econometrician
aims at uncovering the relationship between the same variable at two distant
times in history. The instrument in the historical past identifies the
persistence of a relevant variable, which is, in a sense, the objective
of the analysis; see \reftext{Fig.~\ref{fig:EmpirMod2}}(b). As an illustration of
this type of models we will consider four other classic well-known papers
providing evidence for the persistence of some form of either institutions
or cultural traits over the long-run historical past:
\cite{voi-voth-2012}, \cite{nunn2011slave}, \cite{guiso-sap-zin-2016},
and \cite{alesina-giuliano-nunn-2013}, studying, respectively, the persistence
of anti-semitism, social trust, civic capital, and gender attitudes.

\section{Persistence studies that analyze relationships between current variables}
\label{sec:current}

Consider the special case of the empirical model of Section~\ref{em} in
which the econometrician does not observe any proxy $p_{t-h}$ for
$x_{t-h}$ and $z_{t-h}$ is an instrument for $x_{t}$, through the persistence
of the process $\{x_{\tau}\}_{\tau \in T}$. We assume that $z_{t-h}$ is
a valid instrument to concentrate our analysis on the interpretation of
the estimates of $\beta $ or $\rho $ when they are allowed to be heterogeneous
across locations $i=1,2, \ldots ,N$. Besides requiring that
$z_{t-h}$ be correlated with $x_{t}$, validity also requires
$z_{t-h}$ to be as good as randomly assigned and to satisfy the exclusion
restriction, that is, $z_{t-h}$ affects $y_{t}$ only through $x_{t}$.\footnote{See
\cite{angrist2008mostly}, ch. 4.4.1 for formal assumptions.}

%
\begin{ass}
\label{ass:currentrel}
$z_{t-h}$ is a valid instrument for $x_{t}$; in particular,
\begin{align*}
cov\left (z_{t-h}, x_{t}\right ) & \neq  0
\\
cov\left (z_{t-h}, y_{t} \mid x_{t} \right )&= 0.
\end{align*}
\end{ass}

In \cite{acemoglu2001colonial}, for instance, the process
$\{x_{\tau}\}_{\tau \in T}$ represents the quality of institutions, from
colonial times $t-h$ to the present $t$. The instrument $z_{t-h}$ is settler's
mortality: the authors argue that low settlers' mortality facilitated the
set-up of inclusive institutions by colonial powers, whereas high settlers'
mortality caused extractive institutions. The parameter $\beta $ represents
the returns to institutional quality in terms of economic development,
measured by current per capita GDP. In \cite{ashraf-galor-2013} the process
$\{x_{\tau}\}_{\tau \in T}$ is human genetic diversity, from the first
migration of \emph{Homo sapiens} ``Out of Africa'' until the present. The
instrument is distance of the current geographical location from East Africa,
which is causal to human genetic diversity as a consequence of the \emph{serial-founder
effect}, whereby genetic diversity is reduced at any successive migration
event. The parameter $\beta $ represents the returns to human genetic diversity
in terms of economic development, measured by population density in 1500.
The underlying hypothesis, confirmed by their empirical analysis, is that
diversity has a hump-shaped effect on development, because it has both
beneficial and detrimental effects by increasing social conflict, but fostering
innovation or creativity. \reftext{Table~\ref{tab:papers_current}} presents concisely
the variables adopted in the main specifications of these studies.\footnote{These
examples illustrate well how heterogeneous treatment effects may play a
role in the analysis. \cite{acemoglu2001colonial}, e.g., interpret their
measure of institutional quality a ``cluster of institutions, including
constraints on government expropriation, independent judiciary, property
rights enforcement, and institutions providing equal access to education
and ensuring civil liberties, that are important to encourage investment
and growth. Expropriation risk is related to all these institutional features.''
This \xch{interpretation allows for}{interpretation allow for} different mechanisms connecting expropriation
risk to economic performance, leading naturally to the possibility of heterogeneous
effects, which depend on which mechanism is activated. This is the case
in \cite{ashraf-galor-2013} as well, where genetic diversity, or lack thereof,
affects development by a combination of creativity, increased cooperation,
trust, socioeconomic order, adaptability, specialization, and so on.}

\newcommand{\chead}[1]{\textbf{#1}}
\begin{table}[t]
\caption{Variables in selected current relationship studies.}%
\label{tab:papers_current}
\scalebox{.93}{
\begin{tabular}{cccc}
\toprule
    \chead{Article}  & \chead{$y_{t}$} & \chead{$x_{t-h} $} & \chead{$z_{t-h}$}\\ \midrule 
    AJR (2001)    & GDP per capita, $t= 1995$ & Property rights & Settlers' mortality\\
    AG (2013)    & Population density, $t= 1500$ & Genetic diversity & Distance from Africa \\
\bottomrule
\end{tabular}
}
\medskip

\caption*{ \normalfont \small Note: \xch{See}{see} \reftext{Fig.~\ref{fig:EmpirMod2}}(a) for relationships between variables. Abbreviations:
AJR: \cite{acemoglu2001colonial}, AG: \cite{ashraf-galor-2013}.}
\end{table}

To simplify the intuition and to present results with minimal algebra (the
argument generalizes without these assumptions), consider an environment
where both the instrument $z_{t-h}$, and the treatment $x_{\tau}$, for
any $\tau \in T$, are binary. Under these assumptions, the instrumental
variable estimator coincides with the Wald estimator, whose population
analog is:
%
\begin{align}
\frac{E(y_{t}|z_{t-h}=1)-E(y_{t}|z_{t-h}=0)}
{E(x_{t}|z_{t-h}=1)-E(x_{t}|z_{t-h}=0)} & =   \beta
\label{Wald}
\end{align}
This representation of the IV estimator has an intuitive interpretation:
assuming the instrument is random with respect to the treatment, the effect
of the instrument on the dependent variable (the \emph{Reduced form}) divided
by the effect of the instrument on the treatment (the \emph{First stage})
uncovers the causal effect of a unit change of the independent variable
on the treatment.

Allowing for $\beta $ to be heterogeneous across locations
$i=1,2, \dots , N$ opens the door for the possibility that the relationship
between the instrument $z_{t-h}$ and the independent variable, the treatment
$x_{t}$, is filtered through the effect of $\beta $. Instrumented locations,
with $z_{i,t-h}=1$, might take treatment or not, $x_{i,t}=1$ or $=0$, depending
on $\beta _{i}$. This is the fundamental factor inducing relevant LATE
effects that are different from ATE effects. Importantly, note that this
does not affect the validity of the instrument.\footnote{Validity is guaranteed
if $x_{t} \mid _{ z_{t-h}=1 } \neq x_{t} \mid _{ z_{t-h}=0 }$ and
$y_{t} \mid _{x_{t}, z_{t-h}=0} = y_{t} \mid _{x_{t}, z_{t-h}=1}$, where
$x_{t} \mid _{ z_{t-h}}$ and $y_{t} \mid _{x_{t}, z_{t-h}}$ indicate, respectively,
the random variable $x_{t}$ conditional to the realization of
$z_{t-h}$ and $y_{t}$ conditional to the realization of
$x_{t}, z_{t-h}$. Weaker conditions in terms of conditional means are sufficient.} To formalize the result we want to focus on, assume
$\beta _{i}$ affects differently the relationship between
$z_{i,t-h}$ and the treatment $x_{i,t}$ across locations
$i=1,2\ldots ,N$. Then we can categorize four conceptually different types
of locations, which we label, following the causal identification literature,
\emph{Always takers, Never takers, Compliers, and Defiers}:
\begin{itemize}
\item[] \emph{Always takers}: all $i$ such that $x_{i,t} =1$ for all values
of $z_{i,t-h}$;
\item[] \emph{Never takers}: all $i$ such that $x_{i,t} =0$ for all values
of $z_{i,t-h}$;
\item[] \emph{Compliers}: all $i$ such that $x_{i,t}=1$ when
$z_{i,t-h}=1$ and $x_{i,t}=0$ when $z_{i,t-h}=0$;
\item[] \emph{Defiers}: all $i$ such that $x_{i,t}=1$ when
$z_{i,t-h}=0$ and $x_{i,t}=0$ when $z_{i,t-h}=1$.
\end{itemize}
The instrumental variables estimator in Eq.~\reftext{(\ref{Wald})} identifies
the LATE effects of the instrument, that is, the effect of the instrument
for the Compliers:

%
\begin{thm}
\label{LATETheorem}
\citep{angrist1995identification} Suppose \reftext{Assumption~\ref{ass:currentrel}} holds and that the treatment induces no Defiers, then
\begin{align*}
&  \frac{E(y_{t}|z_{t-h}=1)-E(y_{t}|z_{t-h}=0)}{E(x_{t}|z_{t-h}=1)-E(x_{t}|z_{t-h}=0)}
\\
 ={} &  E\left ( y_{i,t}\mid _{x_{i,t}=1} - y_{i,t}\mid _{x_{i,t}=0}
\mid i \in \text{ compliers} \right )
\\
 = {}&  E(\beta _{i} \mid i \in \text{ compliers} )
\end{align*}
\end{thm}

This theorem helps qualifying the interpretation of the estimate of
$\beta $ in the empirical implementation of Persistence studies when reasonable
structures can be hypothesized which separate Compliers as a subset of
all the locations and on whether in these structures Compliers can be characterized
by a distinct distribution of $\beta $ with respect to the other types.

\subsection{Abstract models of treatment take-up and persistence}

To identify Complier locations and their characteristics it is generally
useful to model how the mechanisms which induces treatment take-up and
persistence, so that $x_{t}=1$, can be filtered through $\beta $. We illustrate
this generally, by developing four minimal abstract models linking
$z_{t-h}$ to $x_{t}$ via $\beta $. These models represent reduced-form
behavioral-equilibrium relationships, as we already noticed, without explicit
micro-foundations. It is our aim to show how this analysis is especially
relevant in an historical context and how it may guide the development
of empirical strategies to uncover the mechanisms underlying the identified
causal effect. We associate these models narratively to the papers we have
selected as an example in this section.

\subsubsection{Treatment take-up at $t-h$}
\label{treatment_t-h}

Let $b_{i}$ denote the benefits of treatment for location $i$ in historical
times, $t-h$; that is, the benefits of $x_{i,t-h}=1$ relative to
$x_{i,t-h}=0$. Assume $b_{i}$ represents a relevant parameter which characterizes
location $i$, but is not observable to the econometrician. Let
$c(z_{t-h})$ denote the cost of treatment at $t-h$; and let the effect
of the instrument be a reduction in the cost of treatment,
\begin{equation*}
c(z_{i,t-h}=1) < c(z_{i,t-h}=0).
\end{equation*}

In \cite{acemoglu2001colonial} lower settler mortality, which we denote with $z_{t-h}=1$, is
assumed to lower the cost of creating and maintaining inclusive institutions
that protect property rights in the colony. In
\cite{ashraf-galor-2013} the distance from East Africa affects treatment
because a shorter distance facilitates higher genetic diversity being associated
with fewer migration events after the original one ``Out of Africa.''

Consider a behavioral-equilibrium relationship postulating treatment at
$t-h$ to be determined by a simple cutoff condition: each location
$i=1,2, \ldots ,N$ takes-up treatment if its benefit is greater than
the cost:
\begin{equation*}
x_{i,t-h}= \left \{
\begin{array}{ll}
1 & \text{ if } b_{i} \geq c(z_{i,t-h})
\\
0 & \text{ otherwise}
\end{array}
\right . .
\end{equation*}
To focus on the relationship between $z_{t-h}$ and $x_{t}$, assume that
the persistence of the explanatory variable is homogeneous (and perfect);
that is, $\rho _{i}=1$, for all $i=1,2\ldots , N$, and
\begin{equation*}
x_{i,\tau}=x_{i,\tau -1}, \; \; t-h\leq \tau \leq t.
\end{equation*}
Treatment can only occur at $t-h$ and is never undermined in the course
of history. Consequently, locations can be categorized depending on their
characteristic $b_{i}$. \emph{Always takers} are all those locations
$i$ such that $b_{i} \geq c(z_{i,t-h}=0) > c(z_{i,t-h}=1)$; and \emph{Never
takers} are all $i$ such that
$ c(z_{i,t-h}=0) >c(z_{i,t-h}=1) > b_{i}$. Most importantly, the \emph{Compliers}
whose effect is identified by the IV are those locations whose benefits
of treatment $b_{i}$ are higher than the cut-off when treated by the instrument,
but lower when not-treated:
\begin{equation*}
c(z_{i,t-h}=1) \leq b_{i} < c(z_{i,t-h}=0).
\end{equation*}
Finally, there are no \emph{defiers} (as required by \reftext{Theorem~\ref{LATETheorem}}).

According to this abstract behavioral-equilibrium model of treatment, for
LATE effects to be distinct from ATE it is sufficient to hypothesize that
$\beta _{i}$ and $b_{i}$ are correlated, i.e. that the benefits of treatment
$b_{i}$ are at least in part obtained through the effects of treatment
on the relevant dependent variable $y_{i,t}$, $\beta _{i}$. In the context
of the effects of institutional quality on economic development in
\cite{acemoglu2001colonial}, it seems natural to think of the returns of
institutional quality in terms of economic development for a country
$i$, $\beta _{i}$, as a measure of the benefits of treatment $b_{i}$ for
this country. In this context, this simple behavioral-equilibrium model of
institutional formation can be used to explicit conditions, that may in
principle be validated empirically, and refine the interpretation of the
IV estimate of the returns of institutional quality by identifying the
mechanisms leading different locations to take up treatment or not.

Consider the following somewhat extreme case as further illustration.\footnote{We
follow \cite{rosenzweig2000natural} who use a similar approach to interpret
IV estimates in several labor economics studies; see the Appendix for an illustrative example} Assume
for simplicity that there are two types of countries, $l,h$ with
$\beta _{h}>\beta _{l}$, and that their \xch{proportions are}{proportions is}
$\pi _{h},\pi _{l}=1-\pi _{h}$ respectively. Further, assume that
$c(z_{i,t-h})$ satisfies $c(z_{i,t-h}=1)<\beta _{l}<c(z_{i,t-h}=0)<\beta _{h}$. High settlers'
mortality, $z_{t-h}=0$, increases the cost for the colonial power to implement
the inclusive institutions that reduce future expropriation risk. Under
this assumption, this cost is generally lower than the benefit, with the
exception of $l$ countries when settlers' mortality is high, which represent
then the \emph{compliers}.\footnote{All that is needed in general is for
settlers' mortality to affect disproportionately one group of countries.
Note that settlers' mortality only affects economic performance by changing
institutional quality directly and therefore is by assumption a valid instrument.}
In this example, therefore, the Wald estimator identifies
%
\begin{align}
\frac{E(y_{t}|z_{t-h}=1)-E(y_{t}|z_{t-h}=0)}{E(x_{t}|z_{t-h}=1)-E(x_{t}|z_{t-h}=0)}
& =   \nonumber
\\
\frac{\beta _{h}\pi _{h}+\beta _{l}\pi _{l} - \beta _{h}\pi _{h}}
{1-\pi _{h}} & =  \beta _{l}
\label{walddaron}
\end{align}
i.e., the gains for countries with low values of $\beta $, the average
value of $\beta $ among the \emph{compliers}:
\begin{equation*}
E\left (\beta _{i} \mid i: c(z_{i,t-h}=0) >b_{i} \geq c(z_{i,t-h}=1)
\right ) =\beta _{l}.
\end{equation*}
Note however that, with a different cost structure, for example in the
opposite extreme case where high-quality institutions are adopted only
by $h$ countries and only when their settlers' mortality is low, the instrumental
variable analysis would on the contrary identify $\beta _{h}$. We conclude
that this analysis of a simple behavioral-equilibrium model of treatment
take-up, while extremely stylized, suggests the importance of evaluating
various specific dimensions institutional change. It suggests notably the
importance of proposing an empirical strategy that tries to test empirically
the underlying assumptions regarding the benefits from institutional change to better understand what effects are identified. It also provides some
nuance about the interpretations of the contemporaneous effects of institutional
quality, because different ``instruments'' favoring improvements in institutional
quality may operate on a different set of compliers.

Related arguments can be used to illustrate the possible role of heterogeneous
treatment effect in the context of \cite{ashraf-galor-2013}. The interesting
difference, in this case\xch{, consists in}{, consist in} the fact that the treatment (genetic
diversity) is not directly the result of a political economy equilibrium
choice, but is rather the (perhaps unintended) consequence of the migration
process via the \emph{serial founder effect}. This mechanism cannot be directly
conceived as the outcome of an equilibrium model. On the other hand, genetic
diversity can also be thought of as the consequence of evolutionary processes,
for example driven by mating patterns in society. Assortative mating along
any phenotypical trait, notably e.g., along ethnic dimensions, may reduce
genetic diversity by producing ethnic cleavages which might turn into distinct
homogeneous populations.

Consider then a behavioral-equilibrium model of mating patterns along migration
events after the original one ``Out of Africa'' as an illustration. Let
instrumented locations $i$, such that $z_{i,t-h}=1$, be those which are
at the economic development-maximizing level distance from East Africa.\footnote{This
is just an innocuous change-of-variable re-normalization of the IV procedure.}
These locations, by construction, are treated by the \emph{serial founder
effect}: their genetic diversity is $x_{i,t}=1$; that is, genetic diversity
is such that the combination of innovation and creativity, on the one side,
and conflict and trust, on the other, induces the highest economic development.
Suppose there are two types of populations defined by their ethnicities,
culture, etc... Type $h$ has a hump-shaped benefit from diversity, type
$l$ also has hump-shaped benefits, but with smaller effects. Let the benefits
at the peak be denoted $b_{h}$ and $b_{l}$, respectively. Suppose that
mating strategies conducive to genetic diversity $x_{t}=1$ are costly -
say they have a cost $c$ - and assume that $b_{l}<c<b_{h}$. Therefore,
\begin{equation*}
x_{i,t-h}=
\begin{cases}
1 & \text{ if } z_{i, t-h}=1; \text{ or if } z_{i, t-h}=0 \text{ and } b_{i}
\geq c
\\
0 & \text{ otherwise}
\end{cases}
.
\end{equation*}
Consequently, the locations with benefits $b_{h}$ are \emph{Always takers}:
they achieve ``optimal diversity'' regardless of location. Locations with
returns $b_{l}$ are instead \emph{Compliers}; that is, they obtain the optimal
diversity $x_{i,t}=1$ only if moving ``Out of Africa.'' As in
\cite{acemoglu2001colonial}, it seems natural to hypothesize that
$\beta _{i}$ and $b_{i}$ are correlated, that the benefits of treatment
$b_{i}$ are at least in part obtained through the effects of treatment
on the relevant dependent variable $y_{i,t}$, $\beta _{i}$. In this case,
the IV strategy identifies the LATE effect, that is, the average
$\beta _{i}$ of the locations that have a different adoption rule when
$z_{t-h}$ changes value\footnote{The Wald arithmetic is identical to \reftext{(\ref{walddaron})}.
In fact, the running hypothesis in \cite{ashraf-galor-2013}, documented
in the data, is more nuanced. Genetic diversity has a hump-shaped effect
on economic development, \xch{reflecting a}{reflecting the a} combination of beneficial and
detrimental factors: diversity is associated with a higher likelihood of
innovation and creativity, but also increases conflict and decreases trust.
The simple model we have delineated can be extended to an environment in
which the instrument and the treatment take three values for
$x_{t-h}$ and $z_{t-h}$, say $0,1,2$, representing respectively ordered
distances from East Africa and measures of genetic diversity. Under the
assumptions that locations with benefits $b_{h}$ have the evolutionary incentive
to converge to the genetic diversity which induces the highest economic
development, independently of their distance from East Africa, and that
$b_{i}$ and $\beta _{i}$ are correlated, the IV strategy identifies
$\beta _{l}$. In this case, the benefits are identified by calculating the
Wald estimator, comparing pairwise locations. First consider locations
$i$ with $z_{i,t-h}= 0, 1$:
\begin{align*}
 & \frac{E(y_{t}|z_{t-h}=1)-E(y_{t}|z_{t-h}=0)}
{E(x_{t}|z_{t-h}=1)-E(x_{t}|z_{t-h}=0)}
\\
& =  \frac{(\pi _{l} \beta _{l}+ \pi _{h} \beta _{h}) - (\pi _{l} 0 +\pi _{h} \beta _{h})}{ 1-\pi _{h}}
= \beta _{l}
\end{align*}
Similarly, calculate the benefit comparing locations $i$ with
$z_{i,t-h}= 1, 2$:
\begin{align*}
& \frac{E(y_{t}|z_{t-h}=2)-E(y_{t}|z_{t-h}=1)}
{E(x_{t}|z_{t-h}=2)-E(x_{t}|z_{t-h}=1)}
\\
& =  \frac{(\pi _{h} \beta _{h}+ \pi _{l} 0) - (\pi _{l}\beta _{l} +\pi _{h} \beta _{h})}
{\pi _{h}+2\pi _{l} -1} = - \beta _{l}
\end{align*}
Suggesting that under the assumptions made on the cost structure, the strategy
identifies the least pronounced hump-shaped relationship between diversity and development.}:
\begin{equation*}
E\left ( \beta _{i} \mid i: b_{i}= b_{l} \right )=\beta _{l}.
\end{equation*}

Indeed, in the context \cite{ashraf-galor-2013}, as in
\cite{acemoglu2001colonial}, conditions can be assumed (in this case on
the costs and benefits of evolutionary selection through mating patterns)
which give rise to distinct implications with respect to LATE effects.
If for instance evolutionary selection towards the optimal diversity does
not operate unless migration occurs, and if even with migration it occurs
only when there are high incentives to do so, then it is the combined effect
of migration and high incentives that generates development. In this case,
the IV would identify $\beta _{h}$, even if diversity may have low or zero
effects on development for some populations. We conclude that this simple
behavioral-equilibrium model in the context of
\cite{ashraf-galor-2013} suggests the importance of empirically evaluating
different mating patterns, and how the structure of benefits (e.g., from isolating
ethnically) depends on migration, to shed light on the mechanisms linking
genetic diversity to economic development. While the stylized model we
used for illustration abstracts from fertility, fertility patters would
also relate to marriage and migration patters in fundamental ways affecting
the interpretation of the identified causal effects of genetic diversity.

\subsubsection{Treatment take-up at $\tau $}

In this section we leverage even more than in the previous one the particular
role of history in the understanding of LATE effects. Consider the abstract
behavioral-equilibrium models delineated in the previous section but assume
now that locations can enter treatment at any time $\tau $, with
$t-h \leq \tau \leq t$, that is, after the realization of the instrument.
One way to introduce the role of history in the determination of treatment
take-up is to let the cost of treatment depend on the instrument
$z_{t-h}$ as well as on the realization of a stochastic process
$\{s_{\tau}\}_{\tau \in T}$: $c_{\tau}=c(z_{t-h}, s_{\tau})$. The empirical
model, with the addition of the effects of the stochastic process
$\{s_{\tau}\}_{\tau \in T}$ is represented in \reftext{Fig.~\ref{fig:EmpirMod3}}.

\begin{figure}[t]
%
\centering
\begin{tikzpicture}%
[observed/.style={circle,draw=black,fill=white,minimum size=11mm},
          unobs/.style={circle,draw=black,fill=white,minimum size=11mm,dashed},
          timetext/.style={draw=white,fill=white},
          node distance=2cm,
          on grid,
          >=latex
        ]%
          \node[observed,red,fill=none] (x) {$x_{t}$};
          \node[observed,right= of x,red,fill=none] (y) {$y_{t}$};
          \node[unobs,below= of x] (xtau) {$x_{\tau}$};
          \node[unobs,below= of xtau] (xth) {$x_{t-h}$};
          \node[unobs,left= of xtau] (s) {$s_\tau $};
          \node[observed,left= of xth] (z) {$z_{t-h}$};
          \node[observed,right= of xth,white] (p) {$p_{t-h}$};
          \node[timetext,above right=.25cm and 1cm  of x] {${\color{red} \beta}$};
          \draw [<->,red]
            (x) edge (y);
          \draw [->,black];
          \draw [->]
            (xth) edge (xtau)
            (xtau) edge (x)
            (z) edge (xth);
         \draw [dashed,->] (s) -- (xtau);
         \draw [dashed,->]  (z) -- (xtau);
          \node[timetext,right= 10mm of y,anchor=west] (pres) {present};
 		  \node[timetext,right = 30mm of xtau, anchor=west] (past) {past};
 		 \node[timetext,right= 10mm of p,anchor=west] {historical past};
    \end{tikzpicture}
\caption{Current variables empirical model: interacting historical process.}
\caption*{\small \normalfont \textit{Note}: \xch{Circles}{circles} indicate variables observed
by the investigator. Dashed circles indicate unobserved variables. Solid
arrows indicate directions of causality. Double arrows indicate endogeneity
or any other factor preventing the identification of a causal effect (omitted
variables, selection bias, etc...).\ Dashed arrows indicate potential causal
links. Highlighted in \xch{red (gray in print version)}{red} is the relationships of interest.}
\label{fig:EmpirMod3}
\end{figure}

In the context of the effects of institutional quality on economic development,
for instance, it is natural to assume the process
$\{s_{\tau}\}_{\tau \in T}$ capturing the dynamics of relevant cultural
variables interacting with the dynamics of institutions.\footnote{\cite{bisin2017joint}
model these interactions in related contexts.} Importantly, the validity
of the instrument $z_{t-h}$, via the exclusion restrictions, is not hindered
by the correlation of $z_{t-h}$ with the cultural process
$\{s_{\tau}\}_{t-h\leq \tau \leq t}$, as long as culture does not have
a direct effect on economic development, that is, as long as
$s_{\tau}$ affects $y_{t}$ only through
$\{x_{\tau}\}_{t-h\leq \tau \leq t}$. Let again the effect of the instrument
be a reduction in the cost of treatment,
\begin{equation*}
c(z_{i,t-h}=1, s_{i,\tau}) < c(z_{i,t-h}=0, s_{i,\tau}).
\end{equation*}
Maintain the assumption that the benefits of treatment across locations
$i$, $b_{i}$, are correlated with $\beta _{i}$; i.e., that the returns
of institutional quality in terms of economic development are a measure
of the benefits of treatment. Suppose for simplicity a form of perfect
persistence, where treatment is never reversed in the course of history,
\begin{equation*}
x_{i, \tau}=1, \; \tau \in T, \text{ implies } x_{i, \tau '}=1, \;
\tau <\tau '\leq t.
\end{equation*}
Treatment is then still determined by a simple cutoff condition, for each
location $i=1,2, \ldots ,N$:
\begin{equation*}
x_{i,\tau}=
\begin{cases}
1 & b_{i} \geq c(z_{i,t-h}, s_{i,\tau}) \text{ or if } x_{i,\tau -1}=1
\\
0 & \text{otherwise}
\end{cases}
.
\end{equation*}
Treatment could occur at any $t-h<\tau \leq t$. The Compliers whose effect
is identified by the IV are those locations whose benefits of treatment
$b_{i}$ are higher than the cut-off when treated by the instrument, but
lower when not-treated; but whether this is the case might depend in general
on the dynamics of $s_{\tau}$. As in the previous case, since
$\beta _{i}$ and $b_{i}$ are correlated, the IV strategy identifies the
LATE effect, that is, the average $\beta _{i}$ of the compliers. In this
case, however, the instrument $z_{t-h}$ interacts with the process
$\{s_{\tau}\}_{\tau \in T}$. In general, $b_{i}$ and the process
$\{s_{i,\tau}\}_{\tau \in T}$ will be correlated. Interpreting
$\beta $ as a measure of the returns of institutional quality effectively
disregards the effect of the dynamics of $s_{\tau}$, which could represent,
as we have already noted, cultural traits or social capital in the historical
process of institutional change.%
\footnote{In fact, $s_{\tau}$ could also be thought of as a contributing
causal factor, e.g., if a different realization of $s_{i, \tau}$ would
have induced an instrumented location $i$ not to set-up high quality institutions
$x_{i,\tau}=1$. In this case we could say that $x_{\tau}, s_{\tau}$ are
jointly causal to $y_{t}$.\label{jo}} The IV strategy identifies in this case the returns to the institutional
quality of the compliers that are activated by the historical processes.
Different, counterfactual histories may have activated processes with different
returns.

To illustrate the role of the interaction between different processes in
the course of history, consider the analysis of the economic development
of the sample of countries colonized by European powers after 1500 in
\cite{acemoglu2002reversal}. This paper documents how i) colonial powers
developed high-quality institutions disproportionally in initially poorer
countries, (a ``Reversal of Fortune''); ii) the inclusive institutions
developed by colonial powers manifested their effects on economic development
only after the Industrial Revolution in 1800-1900, and not before. 

Consider
the two following possible interpretations of these results.\footnote{See
\citet[in this book]{cantoni2020other} for a related argument.} One, assuming wealth in
1500 as exogenous with respect to economic development, is that historical
poverty causes growth. Another interpretation, considering poverty in 1500
as an instrument for beneficial institutional change, is that inclusive
institutions established by colonies cause economic growth. Both interpretations
are valid in principle as long as they are qualified in terms of their
effect being \emph{local}. Poverty in 1500 acted \emph{locally}, through
institutional change in colonial times. But institutional change in colonial
time acted \emph{locally} on economic development through the Industrial
Revolution.\footnote{Following the logic exposed in Footnote \ref{jo}\xch{, we}{, We}
could reasonably consider institutional quality and the Industrial Revolution
as jointly causal.} Both of these \emph{local} effects in principle have
selected a subset of Compliers whose effect the empirical analysis identifies
and which depend on both institutional quality and the Industrial Revolution.
More specifically, the heterogeneity of the effects could be intrinsic
to the quality of institutions as we argued in Section~\ref{treatment_t-h}. But could also be due to the different nature of industrialization
in time or place.\footnote{In fact, in \cite{acemoglu2002reversal}, the
identifying variation in the regression between post-industrialization
production per capita and the interaction of institutions quality and opportunities
to industrialize is the variation of U.K. industrial output over time,
between 1750 and 1980.} Even if the mechanism generating development from
good institutions had homogeneous effect, heterogeneity could arise from
different returns to industrialization.\footnote{A similar discussion could
pertain, for instance, to the negative relationship between past slave
exports and economic performance within Africa, uncovered by
\cite{nunn2008long}, if this effect appears when the dependent variable,
economic performance, is measured in 1980, but not when measured in 1960,
as suggested by \cite{bottero2013there}. This would indicate the interaction
of the effects of slave trade with a more recent phenomenon, like e.g.,
de-colonization.} 

We conclude that this analysis suggests that the interpretation
of the causal effects in Persistence studies depends in a fundamental manner
not only from the historical process of the treatment variable but also
from any other intervening historical process correlated with treatment.
While generally difficult, this calls for the importance of historical
narratives to identify possibly important intervening processes. This is
exactly what \cite{acemoglu2002reversal} do in their study of the effects
of colonization, isolating the Industrial Revolution as the main intervening
factor in the process of development. The introduction of possible heterogeneous
treatment effects adds a layer of complexity and interest, suggesting the
importance to better identify the relationship between the returns to institutional
change and industrialization and their interaction in the development process.

\subsubsection{Treatment take-up and reversals}

In this section we suggest the existence of circumstances in Persistence
studies when one of \xch{the assumptions of}{the assumption of} \reftext{Theorem~\ref{LATETheorem}} may not hold, namely the
absence of Defiers, and study how this changes the interpretation of the
estimates. In the previous sections we have developed minimal abstract
behavioral-equilibrium models of Treatment take-up that consist of simple
cutoff conditions: each location $i=1,2, \ldots ,N$ is treated if the benefits
$b_{i}$ are greater than the cost of adoption, which is reduced by the
treatment and possibly an interacting process, as in the previous section.
In these examples, as we observed, there are no Defiers: a Defier-location
$i$ would have to be characterized by benefits $b_{i}$ such that
$ c(z_{i,t-h}=0) \leq b_{i} < c(z_{i,t-h}=1)$, which contradicts
$c(z_{i,t-h}=1) < c(z_{i,t-h}=0)$.\footnote{The existence of \emph{Defiers}
is ruled out by assumption in \reftext{Theorem~\ref{LATETheorem}}, a special case
of the \emph{Monotonicity} assumption in
\cite{angrist1995identification}. To be precise, what is required is that
Defiers be simultaneously present with Compliers: Defiers would be Compliers
after reversing the definition of treatment.}

Consider as in the previous subsection the case where treatment can be
adopted at $t-h \leq \tau <t$ and the cutoff cost are also affected by
an interacting process $\{s_{\tau}\}_{\tau \in T}$. Suppose the instrument
$z_{t-h}$ - and possibly the treatment $x_{t-h}$ - affect
$\{s_{i,\tau}\}_{\tau \in T}$. In this case, interestingly, non-monotonic
effects of $z_{t-h}$ are possible, without impinging on the validity of
$z_{t-h}$ as an instrument for $x_{t}$. These relationships in the empirical
model are represented in \reftext{Fig.~\ref{fig:EmpirMod4}}.

\begin{figure}[t]
%
\centering
\begin{tikzpicture}%
[observed/.style={circle,draw=black,fill=white,minimum size=11mm},
          unobs/.style={circle,draw=black,fill=white,minimum size=11mm,dashed},
             timetext/.style={draw=white,fill=white},
          node distance=2cm,
          on grid,
          >=latex
        ]%
          \node[observed,red,fill=none] (x) {$x_{t}$};
          \node[observed,right= of x,red,fill=none] (y) {$y_{t}$};
          \node[unobs,below= of x] (xtau) {$x_{\tau}$};
          \node[unobs,below= of xtau] (xth) {$x_{t-h}$};
          \node[unobs,left= of xtau] (s) {$s_\tau $};
          \node[observed,left= of xth] (z) {$z_{t-h}$};
          \node[observed,right= of xth,white] (p) {$p_{t-h}$};
          \node[timetext,above right=.25cm and 1cm  of x] {${\color{red} \beta}$};

          \draw [<->,red]
            (x) edge (y);
          \draw [->,black];
          \draw [->]
            (xth) edge (xtau)
            (xtau) edge (x)
            (z) edge (xth);
         \draw [dashed,->] (s) -- (xtau);
         \draw [dashed,->]  (z) -- (xtau);
         \draw [dashed,->]  (z) -- (s);
         \draw [dashed,->]  (xth) -- (s);

          \node[timetext,right= 10mm of y,anchor=west] (pres) {present};
 		  \node[timetext,right = 30mm of xtau, anchor=west] (past) {past};
 		 \node[timetext,right= 10mm of p,anchor=west] {historical past};
    \end{tikzpicture}
%
\caption{Current variables empirical model: non-linear interacting historical process.}
\caption*{\small \normalfont Note: \xch{Circles}{circles} indicate variables observed
by the investigator. Dashed circles indicate unobserved variables. Solid
arrows indicate directions of causality. Double arrows indicate endogeneity
or any other factor preventing the identification of a causal effect (omitted
variables, selection bias, etc...).\ Dashed arrows indicate potential causal
links. Highlighted in \xch{red (gray in print version)}{red} are the relationships of interest.}
\label{fig:EmpirMod4}
\end{figure}

For instance, it is possible for the process
$\{s_{i,\tau}\}_{\tau \in T}$ to act selectively on locations with
$z_{i, t-h}=0$, fostering high-quality institutions in a subset of these,
against historical odds. E.g., in the context of
\cite{acemoglu2001colonial}, extractive institutions at $t-h$ could foster,
by a mechanism of substitutability between culture and institutions, the
development of cultural traits which over time lead to a reduction of the
cost of high-quality institutions and hence to treatment even location
with relatively low benefits $b_{i}$.\footnote{See
\cite{bisin2017joint} for a formal discussion of substitutability between
culture and institutions.} In this case,
$b_{i}< c(z_{i,\tau}=1, s_{i, \tau})$ for any $\tau \in T$ but
$b_{i}>c(z_{i,\tau}=0, s_{i, \tau}) $ for some $\tau \in T$.
Keeping our assumption of correlation between $\beta _{i}$ and
$b_{i}$ would induce the estimate $\hat{\beta} $ to weigh both Compliers
and Defiers (but not Always-takers and Never-takers). Let $\pi _{c}$ and
$\pi _{d}$ denote the proportion of Compliers and Defiers in the population.
Then, the estimator identifies:%
%
\begin{align}
& \frac{E(y_{t}|z_{t-h}=1)-E(y_{t}|z_{t-h}=0)}{E(x_{t}|z_{t-h}=1)-E(x_{t}|z_{t-h}=0)}
\nonumber
\\[6pt]
={} &E(\beta _{i} \mid i \in \text{ Compliers} )
\frac{\pi _{c}}{\pi _{c}-\pi _{d}} + E(\beta _{i} \mid i \in
\text{ Defiers} ) \frac{-\pi _{d}}{\pi _{c}-\pi _{d}}.
\label{eq:withdefiers}
\end{align}
Expression \reftext{(\ref{eq:withdefiers})} is a weighted average of the treatment
effects, but note that one of the two weights must be negative, therefore
the procedure estimates not a return, but a net effect which is difficult to interpret.\footnote{See \cite{heckman-handbook} for a discussion of the implications of violations
to the monotonicity assumption. \cite{heckman-vytlacil-2005} show in a
general framework that the IV identifies a weighted average of the treatment
effects. In the absence of Monotonicity the weights may be negative. Eq.~\reftext{(\ref{eq:withdefiers})} characterizes this average in the context of our simple
model.}

Examples of the possible role of Defiers in historical contexts typically
include cases of historical reversals in development. Consider
\cite{acemoglu2002reversal} once again, for instance. Relative wealth in
1500 might have fostered future growth, in and of itself, but relative
poverty as well, through institutional change in colonial times and the
Industrial revolution. In this interpretation, countries which have maintained
growth from 1500 would be Compliers, while the protagonists of the \emph{reversal
of fortunes} would be Defiers. Another interesting case of reversal is
documented in \cite{ashraf2010isolation}, when it is shown that more isolated
locations in the Paleolithic are more developed today. It is argued that
isolation was more conducive to innovation, through a mechanism combining
less free-riding on inventions elsewhere and lower predisposition to being
invaded. More generally, we conclude that the interpretation of causal
effects in Persistence studies requires a careful analysis of the distribution
of treatment take-up over historical time, once again suggesting an important
role for historical narratives to guide the formal econometric IV strategy.

\subsubsection{Persistence of treatment}
\label{sec:het_persist}

In this subsection we entertain the analysis of the role of heterogeneous
treatment effects operating through the persistence mechanisms itself rather
than directly through treatment take-up as in the previous sections. Consider
a behavioral-equilibrium model in which treatment at $t-h$ is perfectly
determined by the instrument, $x_{t-h}=z_{t-h}$, but persistence is heterogeneous
and depends on the value of $x_{\tau}$: for instance treatment is more
persistent than lack of treatment. Continuing our parallel with colonial
origins in \cite{acemoglu2001colonial}, this could occur for example because
inclusive institutions, for example, are less costly to maintain once established.\footnote{See
\cite{przeworski2004last} for evidence regarding the heterogeneity of institutional
persistence over historical times.} In the context of (our interpretation
of) \cite{ashraf-galor-2013}, where the persistence of human genetic diversity
is the result of the composition of mating patterns and the serial founder
effect, heterogeneity of persistence could be the consequence of mating
patterns depending on the number of migration events, e.g., because of
the resulting distribution of phenotypical diversity, e.g., ethnic fractionalization.

A simple formalization of heterogeneous treatment effects operating through
the persistence mechanisms is obtained if the transition matrix
$\Pr \left (x_{\tau} \mid x_{\tau -1} \right )$ depends on $\beta $. This
could induce different compliance across values of $\beta _{i}$. To illustrate,
assume $\beta _{i}$ only takes two values, $\beta _{1}<\beta _{2}$, respectively
in fractions $\pi _{1},\pi _{2}=1-\pi _{1}$ of locations, and the correlation
between $x_{t},x_{t-h}$ with $\Pr (x_{t}|x_{t-h})$ is as in \reftext{Table~\ref{tab:het_binary}}.
%
\begin{table}[t]
\caption{Heterogeneous persistence with binary $x$: $\Pr (x_{t}|x_{t-h})$.}
\label{tab:het_binary}
%
\centering 
\begin{tabular}{cc|cc|cc}
    &   & \multicolumn{2}{c}{\chead{$\beta _{1}$}}   & \multicolumn{2}{c}{\chead{$\beta _{2}$}}  \\  
                                  & {$x_{t}$} & 0         & 1           & 0         & 1         \\\hline
\multirow{2}{*}{{$x_{t-h}$}} & 0       & $p_{1}$   & $1-p_{1}$   & $p_{2}$   & $1-p_{2}$ \\
                               & 1       & $1-q_{1}$ & $q_{1}$     & $1-q_{2}$ & $q_{2}$   \\
\end{tabular}
%
\end{table}

The population analog of the Wald estimator is:
\begin{align*}
&
\frac{E(y_{t}|x_{t-h}=1)-E(y_{t}|x_{t-h}=0)}
{E(x_{t}|x_{t-h}=1)-E(x_{t}|x_{t-h}=0)} =
\\[6pt]
&
\frac{\left (
\beta _{1}\pi _{1} q_{1} + \beta _{2}\pi _{2}q_{2}
\right )-
\left (
\beta _{1}\pi _{1} (1-p_{1}) + \beta _{2}\pi _{2}(1-p_{2})
\right )
}
{\pi _{1} q_{1} + \pi _{2} q_{2} - \left ( \pi _{1} (1-p_{1}) + \pi _{2}(1-p_{2}) \right )}
=
\\[6pt]
&
\frac{\sum _{i=1,2}\pi _{i}\beta _{i} (p_{i}+q_{i}-1)}
{\sum _{i=1,2}\pi _{i} (p_{i}+q_{i}-1)}.
\end{align*}
It follows that, as long as $p_{1}+p_{2} \neq q_{1}+q_{2}$, the LATE
effect is different from the ATE. In particular, consider the (natural)
case in which if the persistence of the treatment is correlated with the
returns of the treatment $\beta $, that is,
$\Pr \left (x_{\tau}=1 \mid x_{\tau -1}=1 \right )$ increases with
$\beta $. In the example of the table this would be the case if, e.g.,
$q_{2}>q_{1}$ and $p_{2}=p_{1}$. In this case, the Wald estimator would
identify a LATE effect $\beta $ greater than the ATE effect,
$\sum _{i=1,2}\pi _{i}\beta _{i}$.\footnote{If $p_{1}=p_{2}=1/2$ the LATE
effect would be $\beta _{2}$.} We conclude that a formal behavioral-equilibrium
model of persistence suggests dimensions and directions of inquiry useful
to better qualify the mechanisms driving the causal relationships identified
in Persistence studies that rely on the persistence of the treatment effect
in history.

\section{Pure persistence studies}
\label{sec:purepersistence}

In this section we consider the special case of the empirical model in
Section~\ref{em} in which the objective of the analysis is to prove the
persistence over time of a variable $x_\tau $. In these studies, the econometrician
observes a proxy $p_{t-h}$ for $x_{t-h}$, and, possibly, $z_{ t-h}$, an
instrument for $p_{t-h}$. This empirical model is illustrated in \reftext{Fig.~\ref{fig:EmpirMod2}}(b) which is reproduced here for convenience, as \reftext{Fig.~\ref{fig:EmpirMod5}}. We assume that $z_{t-h}$ is a valid instrument for
$p_{t-h}$ to concentrate our analysis on the interpretation of the estimates
of $\rho $ when they are allowed to be heterogeneous across locations
$i=1,2, \ldots ,N$. Besides requiring that $z_{t-h}$ be correlated with
$p_{t-h}$, validity also requires $z_{t-h}$ to be as good as randomly assigned
and to satisfy the exclusion restriction, that is, $z_{t-h}$ affects
$x_{t}$ only through $p_{t-h}$.
%
\begin{ass}
\label{ass:purepers}
$z_{t-h}$ is a valid instrument for $p_{t-h}$; in particular,
\begin{align*}
cov\left (z_{t-h}, p_{t-h}\right ) & \neq  0
\\
cov\left (z_{t-h}, x_{t} \mid p_{t-h} \right ) &= 0 .
\end{align*}
\end{ass}

The examples of empirical studies that represent this model and we use
as an illustration are listed in \reftext{Table~\ref{tab:purepers}}. All of these
papers are interested in demonstrating the persistence in history of important
factors related to current socio-economic performance: anti-semitism (\cite{voi-voth-2012}),
the role of women in society \cite{alesina-giuliano-nunn-2013}, civic trust
(\cite{nunn2011slave}), and social capital \cite{guiso-sap-zin-2016}.

\begin{figure}[t]
\centering
\begin{tikzpicture}%
[observed/.style={circle,draw=black,fill=white,minimum size=11mm},
              unobs/.style={circle,draw=black,fill=white,minimum size=11mm,dashed},
              timetext/.style={draw=white,fill=white},
              node distance=2cm,
              on grid,
              >=latex
            ]%
              \node[observed,red,fill=none] (x) {$x_{t}$};
              \node[unobs,below= of x,red,fill=none] (xtau) {$x_{\tau}$};
              \node[unobs,below= of xtau,red,fill=none] (xth) {$x_{t-h}$};
              \node[observed,left= of xth] (z) {$z_{t-h}$};
              \node[observed,right= of xth] (p) {$p_{t-h}$};
          \node[above left=1cm and .25cm  of xtau] {${\color{red} \rho}$};
          \node[below left=1cm and .25cm  of xtau] {${\color{red} \rho}$};

            \draw [->,red]
                (xth) edge (xtau)
                (xtau) edge (x);
            \draw [dashed,->]
               (z) edge (xth);
              \draw [<->] (p) edge (xth);
             \draw [->] (z) .. controls +(down:15mm) and +(down:15mm) ..  (p);
          \node[timetext,right= 10mm of y,anchor=west] (pres) {present};
 		  \node[timetext,right = 30mm of xtau, anchor=west] (past) {past};
 		 \node[timetext,right= 10mm of p,anchor=west] {historical past};
        \end{tikzpicture}
%
\caption{Pure persistence empirical model.}
\caption*{\small \normalfont \textit{Note}: \xch{Circles}{circles} indicate variables observed
by the investigator. Dashed circles indicate unobserved variables. Solid
arrows indicate directions of causality. Double arrows indicate endogeneity
or any other factor preventing the identification of a causal effect (omitted
variables, selection bias, etc...).\ Dashed arrows indicate potential causal
links. Highlighted in \xch{red (gray in print version)}{red} is the relationships of interest.}
\label{fig:EmpirMod5}
\end{figure}

While it is possible to find or collect detailed measures of these variables
in current times, the values of these variables in the past can only be
measured with proxies. In addition, in some cases, the proxies are endogenous
to the process and an historical instrument is introduced to disentangle
the causal effect.

\begin{table}[t]
\caption{Variables in selected Pure persistence studies.}%
\label{tab:purepers}
\centering
\begin{tabular}{c|c|c|c}
	\toprule
    \chead{Article}  & \chead{$x_{t}$} & \chead{$ p_{t-h}$} & \chead{$z_{t-h}$}\\ \midrule
    VV (2012)    & \xch{Anti-semitism}{Antisemitism} & Pogroms in 1349  & - \\
    AGN (2013)  & Gender attitudes &  Plow use & Plow suitability \\
    NW (2011)   & Trust  & Slave trade & Distance from  sea \\
    GSZ (2016)   & Civic capital  & City-state & Bishop city \\ \bottomrule
\end{tabular}
\medskip

\caption*{ \normalfont \small Note: See \reftext{Fig.~\ref{fig:EmpirMod2}}(b) for relationships between variables. Abbreviations:
VV: \cite{voi-voth-2012}, AGN: \cite{alesina-giuliano-nunn-2013}, NW:
\cite{nunn2011slave}, GSZ: \cite{guiso-sap-zin-2016}.}
\end{table}

Consider for simplicity (but the arguments extend) an environment where
both $z_{t-h}$, and $p_{t-h}$, are binary and assume that $z_{t-h}$ is
a valid instrument for $p_{t-h}$.\footnote{The validity of the instrument
assumption is formally guaranteed by the following assumption:
\begin{equation*}
\begin{array}{lll}
E\left (p_{t-h} \mid z_{t-h}=1 \right ) \neq E\left ( p_{t-h} \mid z_{t-h}=0
\right ) & & \text{First stage}
\\
x_{t} \mid _{p_{t}, z_{t-h}=0} = x_{t} \mid _{p_{t}, z_{t-h}=1} & &
\text{Exclusion restrictions}.
\end{array}
\end{equation*}} The Instrumental Variable estimator coincides with a Wald estimator:
%
\begin{align}
\frac{E(x_{t}|z_{t-h}=1)-E(x_{t}|z_{t-h}=0)}
{E(p_{t-h}|z_{t-h}=1)-E(p_{t-h}|z_{t-h}=0)} & =   \rho .
\end{align}
Analogously to the cases in Section~\ref{sec:current}, we distinguish:
\begin{itemize}
\item[] \emph{Always takers}: All $i$ such that $p_{i,t-h}=1$ for all values
of $z_{i,t-h}$;
\item[] \emph{Never takers}: All $i$ such that $p_{i,t-h}=0$ for all values
of $z_{i,t-h}$;
\item[] \emph{Compliers}: All $i$ such that $p_{i,t-h}=1$ when
$z_{i,t-h}=1$ and $p_{i,t-h}=0$ when $z_{i,t-h}=0$;
\item[] \emph{Defiers}: All $i$ such that $p_{i,t-h}=1$ when
$z_{i,t-h}=0$ and $p_{i,t-h}=0$ when $z_{i,t-h}=1$.
\end{itemize}
A version of \reftext{Theorem~\ref{LATETheorem}} can be stated in this case.
%
\begin{thm}
\label{LATETheorempers}
\citep{angrist1995identification} Suppose that \reftext{Assumption~\ref{ass:purepers}} holds and that the treatment induces no Defiers, then
\begin{align*}
& \frac{E(x_{t}|z_{t-h}=1)-E(x_{t}|z_{t-h}=0)}
{E(p_{t-h}|z_{t-h}=1)-E(p_{t-h}|z_{t-h}=0)}
\\
 ={}&   E\left ( x_{i,t}\mid _{p_{i,t-h}=1} - x_{i,t}\mid _{p_{i,t-h}=0}
\mid i \in \text{ compliers} \right )
\\
 = {}&  E(\rho _{i} \mid i \in \text{ compliers} )
\end{align*}
\end{thm}
The estimated parameter is generally different from $E(\rho _{i})$ when
persistence $\rho _{i}$ is heterogeneous across locations. We analyze
two special cases.

\subsection{Abstract models of persistence of treatment}

In this section we follow the analysis of Section~\ref{sec:het_persist}, studying in some detail the role of heterogeneous
treatment effects operating through persistence mechanism. We develop behavioral-equilibrium
models in which the persistence of the treatment at $t-h$ is heterogeneous.
We distinguish environments with and without a valid observed instrument
$z_{t-h}$.

\subsubsection{No instrument}

Consider first the simplest case in which the econometrician does not observe
any instrument in historical time $z_{t-h}$. Being interested in documenting
persistence $\rho $ in the process $\{x_{\tau}\}_{\tau \in T}$, the econometrician
only observes $x_{t}$ and a proxy $p_{t-h}$ for $x_{t-h}$. Suppose for
simplicity the underlying process for $x_{t}$ takes the form
\begin{equation*}
x_{t}=\rho x_{t-h}+\epsilon ,
\end{equation*}
where $\epsilon $ is a random shock.\footnote{Recall that $\rho x_{t-h}=\left[\rho_i x_{i,t-h} \right]_{i=1}^N$. } \cite{voi-voth-2012}'s investigation
of the historical persistence of anti-semitism in German cities serves
well as an illustration of this type of Persistence studies. Lacking detailed
evidence of anti-semitism in history, this article uses pogroms in 1348-50CE
in a cross-section of cities as an imperfect proxy for it. The presumption
is that the Black Death of 1348-50CE lowered the threshold for violence
which resulted in Jews being blamed and in some cases being mass-executed
in pogroms as a result. The study finds that pogroms in 1348-50CE are indeed
positively correlated with various detailed measures of anti-semitism in
the 20th century in the cross-section of cities (vote shares for the Nazi
Party, number of deportees from each city, anti-semitic letters to newspapers, etc\ldots ).
In this context, the shock $\epsilon $ captures all determinants of anti-semitism
not operating through its cultural persistence.

Consider a behavioral-equilibrium cut-off model of the relationship between
proxy $p_{t-h}$ and treatment $x_{t-h}$:
%
\begin{equation}
p_{t-h} =
\begin{cases}
1 & \text{ if } x_{t-h} >\widetilde{x}
\\
0 & \text{ if } x_{t-h} \leq \widetilde{x}
\end{cases}
;
\label{pog}
\end{equation}
that is, activation of the proxy is associated to treatment being sufficiently
high, even though the relationship is not necessarily causal. In the context
of \cite{voi-voth-2012}, indeed following their logic, pogroms
$p_{t-h}=1$ occur in cities with high anti-semitism $x_{t-h}$. But it is
not excluded that pogroms might have induced a reinforcement of anti-semitic
attitudes. Lacking data on $x_{t-h}$ it is only possible to estimate
%
\begin{equation}
x_{t}=\alpha ^{R}+\rho ^{R}p_{t-h},
\label{pro}
\end{equation}
which identifies
%
\begin{align}
\rho ^{R} & =   E(x_{t}|p_{t-h}=1)-E(x_{t}|p_{t-h}=0)
\nonumber
\\
& =   \rho \cdot \left ( E(x_{t-h}|x_{t-h}>\widetilde{x})-E(x_{t-h}|x_{t-h}
\leq \widetilde{x}) \right );
\label{rhoRwithOneBeta}
\end{align}
that is, the true persistence parameter $\rho $ times the difference in
treatment between locations with $p_{t-h}=1$ and $p_{t-h}=0$. This kind
of empirical analysis therefore reaches its objective of identifying the
presence of long-run persistence: a positive estimate of $\rho ^{R}$ is
obtained only if $\rho >0$. As long as pogroms are positively correlated
with \xch{anti-semitism}{antisemitism} $x_{t-h}$, using the pogroms variable produces an estimate
of $\rho ^{R}$ that converges, in limit probability, to a positive number.
The estimated $\rho ^{R}$ can be interpreted as the difference between
treated and non-treated locations on the average level of $x_{t}$ which
is due to the persistence effect of $x_{t-h}$.

This interpretation of the estimate of $\rho ^{R}$ can be refined if we
allow persistence to be heterogeneous; that is, if we allow
$\rho _{i}$ to differ across locations $i=1,2, \ldots , N$. As in the examples
from the previous section, heterogeneity may occur because persistence
is somehow correlated with the occurrence of pogroms, or because some other
variable $s_\tau $ affects persistence itself over time. Allowing for heterogeneity
in $\rho _{i}$ across locations, estimating \reftext{(\ref{pro})} identifies
%
\begin{equation}
\rho ^{R} = E(\rho _{i} x_{t-h} | x_{t-h}>\widetilde{x}) - E(\rho _{i}
x_{t-h} | x_{t-h}\leq \widetilde{x}).
\label{rhoRhetBeta}
\end{equation}

An abstract behavioral-equilibrium model of the inter-generational transmission
of cultural traits is helpful in interpreting \reftext{(\ref{rhoRhetBeta})} to illustrate
the implications of heterogeneous persistence. Suppose cultural traits
survive when they are sufficiently strongly held in the population; more
specifically, e.g., a fraction of locations $i$, $\pi _{h}$ with
$x_{t-h}>\overline{x}$ have a high $\rho _{i}=\rho _{h}$ so that the trait
persists more easily over time than in the remaining $(1-\pi _{h})$ cities,
with persistence $\rho _{l}<\rho _{h}$ (see \reftext{Fig.~\ref{fig:pogroms}}).\footnote{See
\cite{bisin2001economics,bisin2011economics} for models of inter-generational
cultural transmission with implications along these lines.}
%
\begin{figure}[t]
%
\centering
\begin{tikzpicture}%
[rect/.style={rectangle,draw=black,minimum height=.6cm},on grid]
            \draw (-2,0.2) -- (5,0.2);
            \draw (-2,0) -- (-2,.4);
            \draw (5,0) -- (5,.4);
            \draw (3,.1) -- (3,.3);
            \draw (2,.1) -- (2,.3);
            \node at(-2.6,.2) {$x_{t-h}$};
            \node at(3,.6) {$\overline{x}$};
            \node at(2,.6) {$\widetilde{x}$};
            \node [rect,minimum width=2cm] at(4,-.4) {$\pi _{h}$};
            \node [rect,minimum width=1cm] at(2.5,-.4) {$\pi _{z}$};
            \node [rect,minimum width=4cm] (pi) at(0,-.4) {$1-\pi _{h}-\pi _{z}$};
            \node [rect,minimum width=2cm] at (4,-1) {$\rho _{h}$};
            \node [rect,minimum width=5cm] (rho) at (0.5,-1) {$\rho _{l}$};
            \node [rect,minimum width=3cm] at (3.5,-1.6) {pogroms};
            \node [rect,minimum width=4cm] (z) at (0,-1.6) {no pogroms};
            \node [left= 2.5cm of pi] {size};
            \node [left= 2.6cm of rho,anchor=east] {$\rho _{i}$};
            \node [left= 2.6cm of z] { $p_{t-h}$};
    \end{tikzpicture}
%
\caption{A model of \cite{voi-voth-2012} with heterogeneous persistence.}
\label{fig:pogroms}
\end{figure}
In this case \reftext{(\ref{rhoRhetBeta})} reduces to
%
\begin{equation}
\rho ^{R} = \rho _{l} E(\widetilde{x} < x_{t-h} \leq \overline{x}) +
\rho _{h} E(x_{t-h}>\overline{x}) - \rho _{l} E(x_{t-h}<\widetilde{x})
\end{equation}
It follows then that the identified effect of persistence
$\rho ^{R}$ can still be interpreted as a difference between treated
and non-treated cities of the average level of $x_{t}$ due to the persistence
of $x_{t-h}$. On the other hand, the size of the effect depends on the
distribution of $x_{t-h}$ and on the relationship between $x_{t-h}$ and
persistence $\rho $ across locations. For instance, assuming instead 
$\widetilde{x}>\overline{x}$, we obtain
\begin{equation*}
\rho ^{R} = \rho _{h} E(x_{t-h}>\widetilde{x}) - \left ( \rho _{l} E(x_{t-h}<
\overline{x}) + \rho _{h} E(\overline{x} < x_{t-h} \leq \widetilde{x})
\right )
\end{equation*}
Under different hypotheses, which, e.g., in \cite{voi-voth-2012}'s context
can ultimately be reduced to how high anti-semitism in the XIVth century
needed to be to trigger a pogrom, the size of the identified effect changes.
Pogroms identify the persistence of anti-semitism in cities where they
took place, not the persistence in cities where they did not occur, because
they did not occur in a random sample of cities.\footnote{This model's assumptions
can be mapped into the assumptions we used to build the model in Subsection \ref{sec:het_persist}. In that model, the persistent variable is binary,
but the correlation between variables over time is more general than the
correlations that are possible imposing the structure of \reftext{Fig.~\ref{fig:pogroms}}. For example, in the model in this subsection, assume
there is an underlying continuous latent variable $\chi _{t-h}$ that determines
both $p_{t-h}$ and $x_{t-h}\in \{0,1\}$ according to different thresholds.
Then, knowing the distribution of $\chi _{t-h}$ one can derive the conditional
probabilities that are analog to $\{p_{i},q_{i}\}, i=1,2$ in \reftext{Table~\ref{tab:het_binary}} and compute what magnitudes the Wald estimator identifies.} We conclude that while it might be difficult to distinguish a priori
between the different process we have hypothesized drive persistence of
anti-semitism in relations to pogroms, it would be interesting in principle
and with additional historical evidence we may be able to support one case
over the other to improve our understanding of the nature of persistence.
More generally, the simple behavioral-equilibrium model indicates novel
and interesting empirical directions to pursue the analysis of the persistence
of cultural traits and values.

\subsubsection{Instrumenting for persistence}

Consider now the case in which the econometrician does observe an instrument
in historical time $z_{t-h}$ for $p_{t-h}$. That is, consider the case
where $z_{t-h}$ affects causally $x_{t-h} $ and $p_{t-h}$, while
$x_{t-h} $ and $p_{t-h}$ are in principle linked by two-way causation,
as in the empirical model in \reftext{Fig.~\ref{fig:EmpirMod2}}(b). The econometrician,
interested in measuring the process' persistence $\rho $, observes
$x_{t}$ and a proxy $p_{t-h}$ for $x_{t-h}$ (but not $x_{t-h}$ directly).

Three examples from the literature clarify this empirical model.
\cite{nunn2011slave} study the persistence of social trust, an important
determinant of current development, using data from African regions. Modern
surveys, such as the Afrobarometer, provide measures of trust in the present.
The historical proxy is the size of slave trade, which varies by location
and ethnicity, and may have affected local trust in the past by generating
``an environment of ubiquitous insecurity caused individuals to turn on
others''. Reverse causality may also hold, however, because communities
with lower trust may experience a lower cost to ``kidnap, trick or
sell each other to slave''). To address this endogeneity the geographical
distance from the coast is used as an instrument for the size of trade.
Distance from the coast correlates with the size of slave trade and, arguably,
does not affect directly current levels of trust. The study finds that
the level of social trust can be traced back to the slave trade.

In \cite{guiso-sap-zin-2016} the process $\{x_{\tau}\}_{\tau \in T}$ represents
civic capital in Italian cities, measured, in modern times, with indicators
such as the proportion of student cheating in mathematics tests, the prevalence
of non-profit associations, or measures of blood donations. The proxy for
this variable in the past is whether the city experienced self-government
in the middle ages, which arguably generated a culture of cooperation and
trust that persists until current times. Indeed, the evidence shows that
cities that experienced self-government have better measures of current
social capital. Because this correlation may be due to omitted endogenous
factors, the paper supports these results by showing that the correlation
remains when instrumenting self-government with the presence of a bishop
seat in the city before 1400 CE. The instrument's validity is motivated 
by the argument (advanced by several historians) that bishops facilitated
the adoption of self-government by morally sanctioning the citizens' agreement.
The presence of bishops lowers the coordination cost required to achieve
independence without affecting modern social capital directly. The study
finds that indeed, self-government in the late middle ages determines higher
civic capital today.

In our last example of papers that conform to this framework,
\cite{alesina-giuliano-nunn-2013}, trace the origin of current cross-cultural
differences regarding the role of women in society (measured by labor-force
participation, political representation, and women entrepreneurship,
$x_{t}$) to agricultural practices in pre-industrial periods, proxied by
plow cultivation ($p_{t-h}$). The process
$\{x_{\tau}\}_{\tau \in T}$ represents gender attitudes and the proxy
$p_{t-h}$ is the adoption of plow cultivation in pre-industrial times.
The argument is that practices requiring more physical strength encouraged
specialization of production by gender, affecting the perception of gender
roles, which persisted in current times even if modern market production
does not require a physical strength advantage. The adoption of plow agriculture
is therefore a proxy for the perception of gender roles, but could be affected
by it, because societies that believe women should be confined to home
production are more prone to adopt agricultural practices that comparatively
advantage men. The study addresses this endogeneity using the suitability
of locations for plow cultivation as an instrument. Plow suitability disproportionally
induces the adoption of the plow, arguably without affecting the perception
of gender roles.

In all these examples, because a valid instrument is observed for the adopted
proxy, the empirical model leads to the identification of
%
\begin{align}
\frac{E(x_{t}|z_{t-h}=1)-E(x_{t}|z_{t-h}=0)}
{E(p_{t-h}|z_{t-h}=1)-E(p_{t-h}|z_{t-h}=0)} & =   \rho .
\end{align}
Following the logic of our discussion of \cite{voi-voth-2012}, an instrument
is not necessary to identify a positive $\rho $. As long as
$p_{i, t-h}=1$ and $x_{i, t-h}$ are correlated, the estimate $\rho $ will
have the correct sign. The endogeneity of the adoption of the plow in
\cite{alesina-giuliano-nunn-2013}, for instance, does not invalidate the
fact that locations which adopted the plow in $t-h$ are characterized by
more conservative gender attitudes at $t$: locations $i$ with
$p_{i, t-h}=1$ have higher $x_{i, t}$. These attitudes in the present are
the consequence of the persistence of cultural traits and attitudes. The
same is true for the identification of the persistence of trust in
\cite{nunn2011slave} and of civic capital in
\cite{guiso-sap-zin-2016}, that we argued have the same structural relationship
between variables.

However, the IV estimate of $\rho $ can be interpreted as a LATE estimate
of the persistence of the process $\{x_{\tau}\}_{\tau \in T}$; that is,
a \emph{local} effect through the specific instrument adopted. For instance,
the IV in \cite{nunn2011slave}, identifies the persistence of community
trust. Assume the benefits of trust $b_{i}$ are heterogeneous across ethnic
groups, for simplicity assume two levels $b_{h}>b_{l}$. Trust is then the
consequence of a behavioral-equilibrium outcome: adopted when its benefits
are greater than the benefits from slave trade. Let the benefit of slave
trade be denoted by $d(z_{t-h})$ and denote with $z_{t-h}=1$ high distance
from the coast, with $d(1)<d(0)$. Assume $b_{l}<d(1)<b_{h}<d(0)$ so that
the ethnic groups with low value of trust $b_{i}=b_{l}$ always experience
slave trade, whereas groups with $b_{i}=b_{h}$ experience trade only if
they are close to the coast. In this case distance from coast induces a
change in treatment only for ethnic groups with the highest returns to
trust, those where, presumably, incentives for trust to persist are higher.\footnote{For
example assume that the cost of acquiring trust is random, but on average
higher than the average cost of maintaining it once acquired. Then ethnic
groups with higher benefits will display higher persistence of trust.}
This would imply that the IV parameter identifies the persistence of
trust in locations where trust is more beneficial and consequently where
its persistence is higher.

Similarly, consider \cite{guiso-sap-zin-2016}, and assume cities have different
returns to civic capital, and persistence is higher when returns are higher.
Further, assume that cities with high returns adopt self-government in
the middle ages regardless of the triggering effect of a coordinating factor
(the presence of a bishop, $z_{t-h}$), whereas cities with
$b_{i}=b_{l}$ adopt self-government only through the coordination from
a bishop (if $z_{t}=1$). In this case Compliers are cities with a low return,
which arguably are less likely to experience persistence otherwise. In
this case, the instrument identifies persistence in cities with relatively
low returns and consequently low persistence.

Finally, in \cite{alesina-giuliano-nunn-2013} the instrument identifies
the persistence of gender attitudes as affected by the exogenous adoption
of the plow. It is conceivable that some locations are Always Takers, that
they have adopted the plow motivated by their previous gender attitudes
- independently of whether their land were suitable for plow agriculture.
If these locations are characterized by stronger persistence of these attitudes,
the instrument identifies persistence in location with relatively lower
persistence.

\section{Conclusions}

In this chapter, we have argued that it is often natural for heterogeneous
effects to occur in the context of Persistence studies: if different mechanisms
affect the variable of interest with different intensity over locations
over time, then different combinations of mechanisms that aggregate into
the same value of the treatment variable generate heterogeneous effects.

Behavioral-equilibrium models can help the interpretation of the causal
effects uncovered in Persistence studies when treatment effects are heterogeneous
and the LATE parameter identified in quasi-experimental designs differs
from the Average Treatment Effect. In this context, we have shown, even
minimal models of treatment take-up may shed some light on what information
the estimated effects provide, at least conditionally on assumptions that
sometimes can be empirically tested. Adding structure, as recent and current
research on the dynamics of institution and culture is doing, can only
be of help; see \citet[in this book]{AceEgor}; \citet[in this book]{BVHHE}; \citet[in this book]{PerssonHHE}. Importantly, these arguments
are independent of the soundness of econometric techniques and of data
reliability.

The interpretation of the estimated parameters is important when Persistence
studies have policy implications. If multiple slow-moving mechanisms (with
heterogeneous effects) underlie long-run correlations, then policies motivated
by the estimated parameters, which are not necessarily the average treatment
effect, may produce unexpected outcomes if they do not operate through
the channels that are \emph{locally} identified by the research design.

The correct interpretation of the estimated parameters is important not
only for policy implications, but also when the researcher is interested
in counterfactual historical analysis. When treatment effects are heterogeneous,
the instrument operates by identifying a specific historical path whose
effects may be different from those that may have been generated in a counterfactual.%

Pushing the arguments of this chapter forward, this line of analysis sheds
doubts on the notion of \emph{origin}, often used in these studies. A causal
relationship realized in history at time $\tau $ with effects at time
$t$ does not preclude another historical causal relationship realized prior
to $\tau $.%
\footnote{For instance, the origin of the Mafia in Sicily has been reduced
with good arguments to the rise of socialist Peasant Fasci organizations
at the end of the 19th century \citep{acemoglu2020weak}; to a price shock
on sulphur and lemon in the 1850's
\citep{buonanno2015poor,dimico2017origins}; to the lack of city states
in the XIV'th century - in turn a consequence of Norman domination
\citep{guiso-sap-zin-2016}; to the Paleolithic split into nomadic pastoralism
in 7th millenium B.C. \citep{alinei2007origini}.} In many cases the relevant
question is not what was the origin of a phenomenon, but what are the counterfactual
\emph{quantitative} effect that results from changing variables in the historical
past on variables in the present. Each variable at different points in
the past may affect the present differently. This paper highlights how
the empirical implications of Persistence studies may be interpreted using
simple models that inform how these mechanisms affect the variable of interest
suggesting directions for future research.\looseness=1


\appendix
\section*{LATE for school}

The clearest applications of the distinction between LATE and ATE are in
labor/education economics. In this Appendix we exploit
\citet{rosenzweig2000natural}'s comment on \citet{angrist1991does}'s instrumental
variable approach to estimating returns to schooling.\footnote{Underlying
this analysis is the debate in econometrics regarding identification power
in reduced-form causal inference design: see
\cite{angrist1991does,angrist1995identification,angrist1999comment,deaton2010instruments,deaton2020,heckman1997instrumental,heckman1999instrumental,heckman2010comparing,imbens2010better}.}

Consider the relationship between schooling attainment and earnings. A
simple regression fails to identify the causal effect of schooling because
an omitted variable bias: for example, children of higher ability may earn
higher wages for given schooling and also choose a higher schooling attainment.
A valid instrument for schooling may be adopted to identify the causal
effect. \cite{angrist1991does} propose to use Quarter of birth. The birth
date cutoffs for school-entry age combined with minimum compulsory schooling
ages induce some children born during the last months of the year to complete
more years of schooling relative to children born at the beginning of the
year, because they are induced to start schooling at an earlier age (this
is true for example if they intend to leave school at the mandatory minimum
age). The arguably random variation in date of birth provides a ``natural''
instrument for estimating the return to schooling.\footnote{In fact, the
validity of quarter of birth as an instrument has been questioned, see
\cite{buckles-2013}, but for the purposes of this introduction, we assume
the instrument to be valid from an econometric standpoint.}

Consider the case in which the treatment effect is heterogeneous, that
is, returns to schooling vary in the cross-section of students; for instance,
they increase with the underlying unobservable ability. To illustrate the
difference between LATE and ATE, limit the students' choice to one extra
year of schooling after mandatory schooling age, and assume that high ability
students attain an extra year of education regardless of their Quarter
of birth, but low-ability students always intend to drop out of school
at mandatory age, and can only be ``forced'' to undertake an extra year
of education if, by being born late in the year, they start attending school
one year younger.

The instrumental variable technique ``works'' by randomly inducing a subset
of students, low-ability students, to undertake an extra year of education.
The extra earnings gained by students not born in the first quarter, relative
to the earnings of those born during the rest of the year, are generated
only by low-ability students that attend an extra year of education. All
other students have the same schooling attainment regardless of the date
of birth, therefore the instrumental variable estimate only identifies
the returns for low-ability students (a LATE effect), not the return of
high-ability students, nor the average return of all students (the ATE).

While the estimated parameter remains of great interest, this interpretation
is consequential in that it suggests that the specific IV procedure adopted,
while valid, under-estimates the returns to schooling. Furthermore, it
suggests caution when considering the policy implications of the estimates,
because a different policy with the goal of inducing higher schooling attainment
may not induce the same subset of students to comply.

This stylized model highlights how the Quarter of birth instrument is likely
to identify the returns of the low-ability students, because arguably these
students are the most likely to leave schooling at the minimum compulsory
age. It is both a logically reasonable assumption, and an empirically testable
implication that could provide additional evidence about the interpretation
of the estimated returns. It is in this spirit that we proceed, in this
paper, to formalize stylized models of persistence studies.

\newpage

\bibliographystyle{econ}
\bibliography{bm-proofs}
\end{document}